\documentclass[%
reprint,
superscriptaddress,
showpacs,preprintnumbers,
amsmath,
amssymb,
aps,
pra,
]{revtex4-1}
\usepackage{graphicx}
\usepackage{dcolumn}
\usepackage{bm}
\usepackage{hyperref}
\usepackage{color}

\def\ul{\hat{U}_{loc}}
\def\lueq{=\ul}

\newcommand{\bra}[1]{\ensuremath{\left\langle#1\right|}}
\newcommand{\ket}[1]{\ensuremath{\left|#1\right\rangle}}

\def\ketbra#1#2{{\ket{#1}}{\bra{#2}}}

\def\BS#1#2#3{\hat{B}_{{#1}{#2}}(#3)}
\def\BSd#1#2#3{\hat{B}^\dagger_{{#1}{#2}}(#3)}

\makeindex

\begin{document}
	\title{Entanglement Enhancement in Multimode Integrated Circuits}
	\author{Zacharie M. L\'{e}ger}
	\email{zacharie.leger@mail.uotronto.ca}
	\affiliation{The Edward S. Rogers Department of Electrical and Computer Engineering, University of Toronto, 10 King’s College Road, Toronto, Ontario M5S 3G4, Canada}
	\author{Aharon Brodutch}
	\affiliation{The Edward S. Rogers Department of Electrical and Computer Engineering, University of Toronto, 10 King’s College Road, Toronto, Ontario M5S 3G4, Canada}
	\affiliation{Department of Physics, University of Toronto, 60 St George St, Toronto, Ontario, M5S 1A7, Canada}
	\affiliation{Center for Quantum Information and Quantum Control, University of Toronto, 60 St George St, Toronto, Ontario, M5S 1A7, Canada}
	\author{Amr S. Helmy} 
	\affiliation{The Edward S. Rogers Department of Electrical and Computer Engineering, University of Toronto, 10 King’s College Road, Toronto, Ontario M5S 3G4, Canada}
	\affiliation{Center for Quantum Information and Quantum Control, University of Toronto, 60 St George St, Toronto, Ontario, M5S 1A7, Canada}
	
	\date{\today}
	
	\begin{abstract}
		The faithful distribution of entanglement in continuous variable systems is essential to many quantum information protocols. As such, entanglement distillation and enhancement schemes are a cornerstone of many applications. The photon subtraction scheme offers enhancement with a relatively simple setup and has been studied in various scenarios. Motivated by recent advances in integrated optics, particularly the ability to build stable multimode interferometers with squeezed input states, a multimodal extension to the enhancement via photon subtraction protocol is studied. States generated with multiple squeezed input states, rather than a single input source, are shown to be more sensitive to the enhancement protocol, leading to increased entanglement at the output. Numerical results show the gain in entanglement is not monotonic with the number of modes or the degree of squeezing in the additional modes. Consequently, the advantage due to having multiple squeezed inputs states can be maximized when the number of modes is still relatively small (e.g., $4$). The requirement for additional squeezing is within the current realm of implementation, making this scheme achievable with present technologies.
	\end{abstract}
	
	\maketitle
	
	\section{Introduction}
	Entanglement is one of the most distinct tenets of quantum mechanics. A plethora of tasks in quantum information science and technology, including teleportation \cite{Teleportation}, distributed quantum computing \cite{NielsenAndChuang, QuantumInternet}, and dense coding \cite{DenseCoding}, necessitate the reliable generation and transmission of entangled states in optical circuits and networks. The efficacy of such multi-party protocols depends on the strength of the quantum correlations in shared entangled states. Unfortunately, highly entangled states of light are difficult to produce. Moreover, loss during generation, transmission, and processing degrades the quality of the entanglement, and thereby the protocol. Quantum error correction and entanglement distillation schemes have been developed and applied to increase a system's resilience to losses and other forms of noise. However, such schemes are usually difficult to implement since they involve many subsystems and can require non-linear photon-photon interactions, active components, and quantum memories \cite{Schnabel}. A subset of protocols can still enhance entanglement without these requirements, at the cost of being sub-optimal. One example proposed by Opatrn\'{y} et al. \cite{OriginalGaussianDistillation} is entanglement enhancement via photon subtraction (EvPS), which only requires a beam-splitter type coupling between modes and high-efficiency single-photon detectors. The relative simplicity of this protocol and the possible advantages it offers in improving the quality of the entangled states have made it the subject of many studies \cite{OriginalGaussianDistillation, SinglePhotonSubDistillation, Fiurasek2011, LocalSq, Navarrete-Benlloch, Walmsley}. 
	
	Recent results have shown a significant boost in the capabilities of integrated quantum photonic devices, where a self-pumped integrated source of entangled photons (e.g., two-mode squeezed states) have already been realized in practical, scalable platforms, thereby rendering large-scale quantum photonic circuits a reality \cite{SelfPump}. Such monolithic device capabilities enable more extensive networks of non-classical light sources to be built in a stable, scalable, and integrated setting. In integrated circuits, present technologies can readily support multimodal interferometer networks, which cannot be stabilized and scaled up to free-space using tabletop approaches. Two critical capabilities that are enabled in multimode devices are the possibility to share entanglement between multiple parties, and the ability to generate multi-mode states that demonstrate higher entanglement than two-party schemes; both of which will be examined here in the context of EvPS.
	
	After introducing the formalism in Sec. \ref{Sec:Formalism}, the EvPS scheme is described in Sec. \ref{sec:single-source} where the two and $N$ mode cases with one single mode squeezed vacuum (SMSV) input followed by an interferometer generate the entanglement. The extension to multiple squeezed vacuum input sources is given in  Sec. \ref{sec:multi-source}, with numerical results and a comparison of the schemes given in Sec. \ref{sec:performance}.  The results are discussed and summarized in Sec. \ref{sec:Discussion}. Throughout, the logarithmic negativity quantifies the entanglement, and the interferometer is set up so that the initial state becomes symmetric for all parties. Since there are no generic methods to quantify multipartite CV entanglement, the approach from Vidal and Werner (2000) \cite{LogNeg}, which studies how the entanglement behaves for all possible partitions (see Appendix \ref{sec:traces}), is adopted. The results show a significant advantage in using multiple sources of squeezed light. Moreover, in some cases the greatest advantage appears when the initial squeezing in the first mode is much higher than in the other modes, suggesting a more experimentally feasible architecture. 
	
	\section{Formalism} \label{Sec:Formalism}
	\subsection{Continuous-variable systems}
	Like many other quantum protocols, entanglement distillation was originally designed for discrete variable systems, usually qubits. This paradigm is not always applicable, as many systems carry continuous degrees of freedom. It is common to associate each mode with the corresponding annihilation operator $\hat a_i$, $i\in\{1,2..,N\}$ and use the Fock basis to denote the state of the system. CV states of light offer efficient preparation, manipulation, and often near-unity efficient detection of entangled states, making them solid candidates for practical implementations using existing technologies \cite{Braunstein2005}.
	
	One subset of operations that can be applied to CV states is the set of Gaussian operations: squeezing, mode mixing (e.g. beam splitting), local phase operations, homodyne detection, partial trace, the addition of a mode in a thermal state and classical communication. In this work, the phase free beam splitter (BS) operation for modes $i$ and $j$, is defined as
	\begin{equation}
	\hat{B}_{ij}(\theta)=\exp\left[\left(\theta-\frac{\pi}{2}\right)\left(-\hat{a}_i\hat{a}_j^\dagger+\hat{a}_i^\dagger\hat{a}_j\right)\right],
	\end{equation}
	while the single mode squeezing operator on mode $k$, $\hat{S}_k(\zeta)$ and the two mode squeezing operator for modes $m$ and $n$, $\hat{S}_{mn}(\zeta)$, are denoted 
	\begin{align}
		\hat{S}_{k}(\zeta)&=\exp\left(\frac{1}{2}(\zeta^{*}\hat{a}_{k}^{2}-\zeta(\hat{a}_{k}^{\dagger})^{2})\right)\\
		\hat{S}_{mn}(\zeta)&=\exp\left(\zeta^{*}\hat{a}_{m}\hat{a}_{n}-\zeta\hat{a}_{m}^{\dagger}\hat{a}_{n}^{\dagger}\right),
	\end{align}
	where $\zeta=re^{i\theta}$ is the squeezing parameter.
	
	Gaussian states are the subset of CV states that have a Gaussian Wigner function or equivalently the set of states that can be generated from the vacuum using Gaussian operations. They are completely characterized by their first and second statistical moments of the quadrature field operators. However, only the second moments, given in terms of a covariance matrix, carry information about correlations and entanglement. As such, the first moments can always be reduced to zero through unitary operations on individual modes \cite{Experiment}. If two Gaussian states have the same covariance matrix up to some local transformations, they are equivalent regarding their entanglement.
	
	The archetypal entangled Gaussian state is the two-mode squeezed vacuum (TMSV) which can be generated, for example, by interfering two SMSV states with an appropriate relative phase at a balanced BS. The TMSV state is a particular case of a more general family of states sometimes referred as CV Greenberg-Horne-Zeilinger (GHZ) states since the sum of the momentum of all output modes and the difference between the position of any two output modes is well characterized; making an allusion to the properties of the discrete variable GHZ state \cite{QIwCV}. Moreover, CV GHZ states are also completely symmetric across all modes, implying mode indistinguishably. Furthermore, every mode and grouping of modes is entangled with every other mode and grouping, making it ideal for many CV quantum information protocols \cite{Braunstein2005,vanLoock2000}.
	
	Entanglement distillation has been an important tool in quantum information, but early distillation protocols relied on states having non-Gaussian Wigner functions \cite{Key15}. However, a no-go theorem established that systems with Gaussian Wigner functions could not be distilled using only Gaussian operations \cite{no-go1,no-go2,CiracGaussian}. Moreover, non-Gaussian states or operations are required for efficient universal quantum computing using CVs \cite{Weedbrook2012}. As such, a number of non-Gaussian transformations have been proposed \cite{OriginalGaussianDistillation,imp2,imp3,imp4,imp5,imp6,Schnabel}. The proposal of Opatrn\'{y} et al., involved subtracting a single photon from each mode of a two-mode squeezed vacuum (TMSV) with the use of low reflectivity BSs \cite{OriginalGaussianDistillation}. However, it was later shown by Ourjoumtsev et al. \cite{SinglePhotonSubDistillation} that having a single photon detected in one mode would also lead to enhancement, but would not keep the Gaussian-like properties of the original state, which are required for tasks like teleportation. In these protocols, enhancement results from a conditional measurement of a single photon in a weakly reflected beam, see Fig. \ref{fig:2mode}, which increases the mean photon number in the transmitted mode (even though photons are subtracted). 
	
	\subsection{Entanglement and gain}
	One computationally simple and relatively standard method to quantify entanglement in bipartite systems is the logarithmic negativity, $E_{\mathcal{N}}(\rho)$, see Appendix \ref{sec:Log-Neg}. As expected from such a measure, the logarithmic negativity is invariant under local unitary operations. This fact will be used to simplify calculations, and $\ul=\prod_k \hat{U}_k$ will indicate a unitary which can be decomposed into a sequence of local unitary operations, $\hat{U}_k$, one for each mode $k$. More generally, $E_{\mathcal{N}}$ is non-increasing under (deterministic) local operations and classical communication. These features make the logarithmic negativity a valid entanglement monotone, i.e., a reasonable metric for quantifying entanglement.
	
	Monotones for multipartite entanglement are not easy to define. However, by examining collections of different bipartite splittings of the system, several inequivalent computational measures of entanglement can be obtained \cite{LogNeg}. Then the multipartite entanglement is said to have increased (decreased) if the logarithmic negativity increased (decreased) for all bipartition \cite{LogNeg, Dur}.
	
	To help contrast the entanglement of a state before and after photon subtraction, $\rho_0$ and $\rho_1$ respectively, the gain in entanglement, $G(\rho_1)$, as defined in \cite{Walmsley} is used,
	\begin{equation}
	G(\rho_1)\equiv\frac{E_{\mathcal{N}}(\rho_1)-E_{\mathcal{N}}(\rho_0)}{E_{\mathcal{N}}(\rho_0)}.
	\end{equation}
	
	\section{Multimode entanglement enhancement with a single source} \label{sec:single-source}
	\subsection{Two modes}\label{subsec:two_modes}
	The EvPS scheme is based on the ability to implement an approximate non-deterministic photon subtraction operation. The detection of a photon, through a weak coupling into a new mode, under the assumption that a detection event only happens in a single mode, can be approximated by an annihilation operator \cite{PhotonSubTheory}. A successful photon subtraction event in one of the modes is triggered by the corresponding detector firing. For simplicity, the photon subtracted mode is labeled as $A$. Note that the symmetric construction (having weak BSs and detectors in all modes) is not an essential feature of the protocol. 
	\begin{figure}[h]
		\includegraphics[width=\columnwidth]{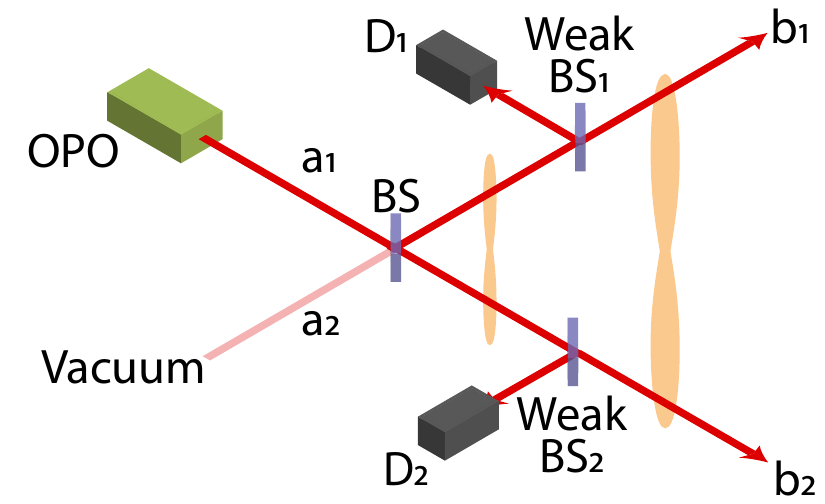}
		\caption{A free-space implementation of the EvPS scheme. A SMSV  is incident on a 50/50 beam splitter (BS), generating a two mode entangled state.  Photon subtraction is conditioned on the detection of a photon at a detector, $D_1$  or $D_2$, after passing through low-reflectivity (weak) BSs. The result is a state which is more entangled than the state after the first BS.}
		\label{fig:2mode}
	\end{figure}
	For an incoming state $\ket{\psi_0}$, a photon subtraction event produces the outgoing state
	\begin{equation}\label{eq:photo-sub}
	\ketbra{\psi_1}{\psi_1}\approx \frac{\hat{b}_A\ketbra{\psi_0}{\psi_0}\hat{b}^\dagger_A}{\sqrt{\mathrm{Tr}(\hat{b}^{\dagger}_A\hat{b}_A\ketbra{\psi_0}{\psi_0})}}.
	\end{equation}
	As an example of the EvPS protocol, the state $\ket{\psi_0}$ prepared by interfering a SMSV state with squeezing parameter $\zeta$ in mode $1$ and the vacuum in mode $2$ (see Fig. \ref{fig:2mode}) using a balanced BS is considered. This state can be prepared relatively easily in a tabletop experiment, and will serve as a benchmark for the multiple input designs. 
	
	Using the result of Eq. \eqref{eq:photo-sub}, the entanglement enhancement can intuitively be understood for a weak initial squeezing, $|\zeta|\ll1$, by noting that 
	\begin{multline}\label{eq:Bell} 
		\ket{\psi_1(2,\zeta)}\equiv\hat{b}_{A}\hat{B}\left(\frac{\pi}{4}\right)\hat{S}_1(\zeta)\ket{0,0}_{1,2}\\
		\approx\hat{b}_{A}\left(\ket{0,0}_{1,2}-\frac{\zeta}{2\sqrt{2}}(\ket{2,0}_{1,2}+\sqrt{2}\ket{1,1}_{1,2}
		+\ket{0,2}_{1,2})\right.\\
		\left.+O(\zeta^{2})\right)
		=\frac{1}{\sqrt{2}}\left(\ket{1,0}_{1,2}+\ket{0,1}_{1,2}\right)+O(\zeta),
	\end{multline}
	where $\hat{b}_{A}$ is the annihilation operator in either one of the output modes (1 or 2).
	The state $\ket{\psi_1(2,\zeta)}$ is a maximally entangled Bell state, up to a correction of $O(\zeta)$, which is an improvement on the degree of entanglement of the original state since it was the vacuum state to zeroth order. It is possible to find an analytical expression for the logarithmic negativity of the exact state $\ket{\psi_1(2,\zeta)}$ in \eqref{eq:Bell} as a function of $\zeta$ and show that the gain in entanglement is always $G(\rho_1)>1$ \cite{Walmsley}. 
	
	\subsection{Multimode}\label{subsec:multimode}
\begin{figure}
		\includegraphics[width=\columnwidth]{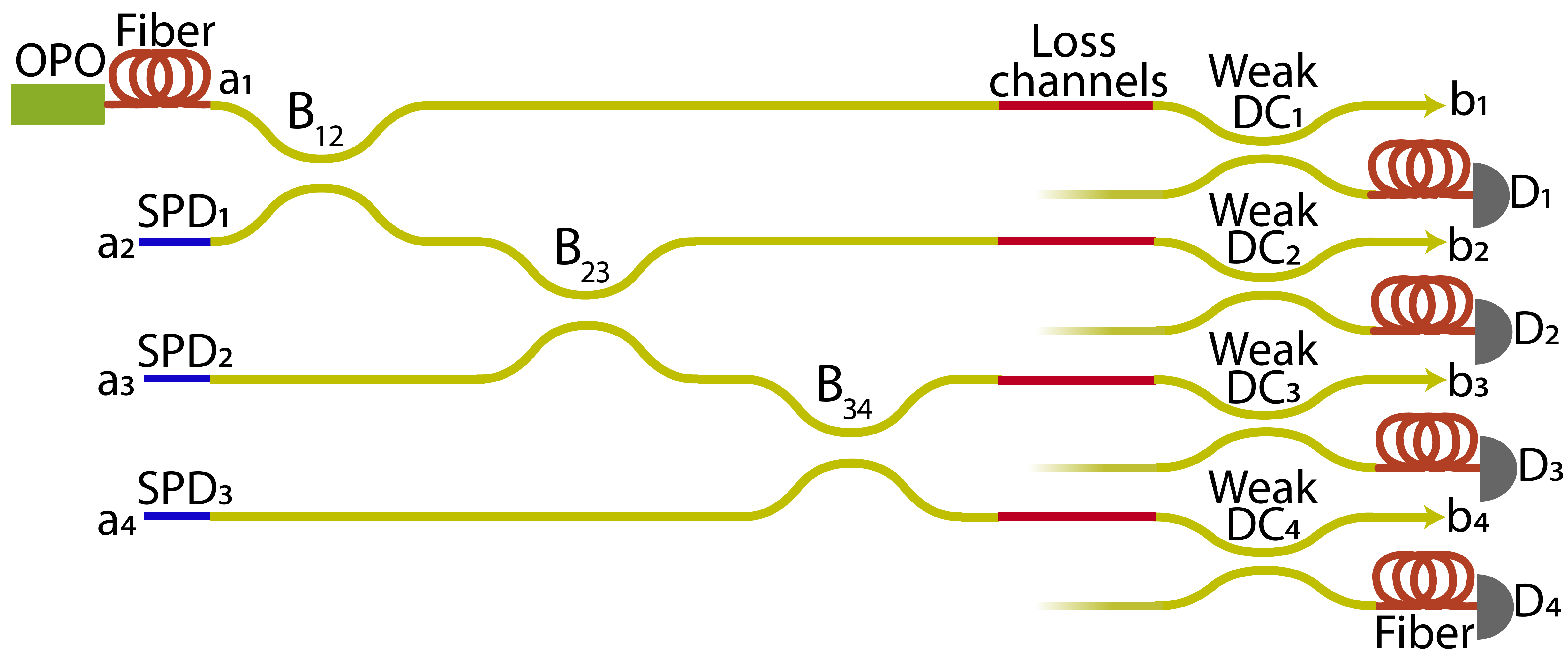}
		\caption{A photon subtraction scheme from an integrated four mode CV GHZ source. A single mode squeezed vacuum (SMSV), $\hat{S}_1(-r_1)\ket{vac}$, is pumped into mode $\hat{a}_1$. Similarly, SMSVs, $\hat{S}_2(r_2)\ket{vac}$, are pumped coherently in modes $\hat{a}_2$, $\hat{a}_3$, and $\hat{a}_4$ through self-pumped devices (SPD). These squeezed states are interfered through directional couplers, $\hat{B}_{12}$, $\hat{B}_{23}$, and $\hat{B}_{34}$, to generate a CV GHZ state.  Photon subtraction is achieved through the detection of a photon in one of detectors, $D_{1}$, $D_{2}$, $D_{3}$, or $D_{4}$, after passing through weak directional couplers, $Weak\,DC_{1-4}$. Note, the photon subtraction operations do not need to be applied in all modes.}
		\label{fig:wgarray}
	\end{figure}

	The scheme above can be generalized to a symmetric $N$ mode state prepared using a single source. In practice, it is possible to prepare the initial state using an on-chip interferometer where directional couplers are used to implement the BS operations \cite{BosonSampling}. 
	For example, a  single-mode squeezed vacuum input in mode $\hat{a}_1$ can be symmetrically split across $N$ modes through a series of directional couplers as seen in Fig. \ref{fig:wgarray}, generating the state 
	\begin{multline}\label{eq:psi}
		\ket{\psi_0(N, \zeta)}=\hat{\mathcal{U}}(N)\hat{S}_{1}(\zeta)\ket{vac}\\
		=\exp\left(\frac{\zeta^*}{2}\left(\frac{\hat{b}_{1}+...+\hat{b}_{N}}{\sqrt{N}}\right)^2-h.c.\right)\ket{vac}_{1...N}.
	\end{multline}
	where $\mathcal{\hat{U}}(N)=\hat{B}_{N-1\,N}\left(\arcsin\frac{1}{\sqrt{2}}\right)...\hat{B}_{12}\left(\arcsin\frac{1}{\sqrt{N}}\right)$ is the $N$-mode phase-free symmetric splitter unitary which is fully described in Appendix \ref{sec:def}.
	
	To compare entanglement measures before and after photon subtraction, the logarithmic negativity (and gain) for all bipartite splittings $\mu$ and $\nu$, with each grouping containing $n$ and $m$ modes respectively (where $n+m=N$) is considered. One can rewrite the initial state as
	\begin{equation}
	\ket{\psi_0{(N,\zeta)}}=\exp\left(\frac{\zeta^*}{2}\left(\frac{\sqrt{n}\hat{b}_{\nu}+\sqrt{m}\hat{b}_{\mu}}{\sqrt{N}}\right)^2-h.c.\right)\ket{vac}_{\nu,\mu}
	\end{equation}
	where
	\begin{equation}\label{eq:split}
	\hat{b}_{\nu}=\frac{\hat{b}_{i_{1}}+\hat{b}_{i_{2}}+...+\hat{b}_{i_n}}{\sqrt{n}},\; 
	\hat{b}_{\mu}=\frac{\hat{b}_{j_1}+\hat{b}_{j_2}+...+\hat{b}_{j_m}}{\sqrt{m}},
	\end{equation}
	here $i_1, ..., i_n$ and $j_1, ..., j_m$ denotes any ordering of the total $N$ modes.
	
	Intuitively, for a low squeezing parameter
	\begin{multline}\label{eq:psismall}
		\ket{\psi_1{(N,\zeta)}}\equiv\hat{b}_A\ket{\psi_0{(N,\zeta)}}\\
		\overset{|\zeta|\ll1}{\approx} \hat{b}_A\left(\ket{vac}_{1,...,N} + \frac{\zeta}{2}\left( \frac{\sum_{i=1}^{N}\hat{b}^\dagger_i}{\sqrt{N}} \right)^2\ket{vac}_{1,...,N}\right)\\
		=\frac{\sum_{i=1}^{N}\hat{b}^\dagger_i}{\sqrt{N}}\ket{vac}_{1,...,N}=\frac{\sqrt{n}\hat{b}^\dagger_\mu+\sqrt{m}b^\dagger_\nu}{\sqrt{N}}\ket{vac},
	\end{multline}
	where $\hat{b}_A$ is the annihilation operator in any of the $N$ output modes. The logarithmic negativity  calculated with respect to this new subsystem is
	\begin{equation}
	E_{\mathcal{N}}(\rho_{\psi_1})=\log_{2}||\rho_{\psi_1}^{T_A}||_1\overset{|\zeta|\ll1}{\approx}\log_{2}(1+2\sqrt{nm}/N),
	\end{equation}
	where the strength of the quantum correlations between the subsystems $\mu$ and $\nu$ can be maximized by a partition where $n=m$ for an even number of modes $N$, alternatively $n=m\pm1$ for an odd number of modes.
	
	This can be seen by looking at the output state of Eq. \eqref{eq:psismall}, where, when $n=m$, perfect anti-correlations are observed in the $\{\ket{0}_i,\ket{1}_i\}$ basis and perfect correlations are also seen in an uncorrelated basis, $\ket{\pm}$, where $\ket{\pm}_i = \frac{\ket{1}_i \pm \ket{0}_i}{\sqrt{2}}$, with $i=\{\mu,\nu\}$.
	
	Alternatively, this could be understood from the von Neumann entropy of the subsystem $\mu$ or $\nu$ of the state \eqref{eq:psismall}, which is maximal when $n=m$, but decays to zero when either $n$ or $m$ tends to $N$.
	
	Note that the logarithmic negativity of $\ket{\psi_1{(N,\zeta)}}$ for larger $\zeta$ can be numerically obtained for any number of modes, $N$, without increased computational power by exploiting the superpositions of modes as shown in \eqref{eq:split}.
	
	\section{Multimode entanglement enhancement with multiple sources} \label{sec:multi-source}
	The ability to use natural non-linearities in integrated waveguides as squeezed vacuum sources opens new opportunities for generating multimode states \cite{SelfPump, Bijlani2013, Aharon}. In particular it is possible to generate CV GHZ states that are much more sensitive to EvPS than those discussed in section \ref{sec:single-source}, the main goal of this sections is to show that this is true and to explore the requirements on these sources. For that purpose, it is sufficient to consider the three parameter subset of states 
	\begin{equation}\label{eq:GHZ}
	\ket{\phi_{0}(N,r_1,r_2)}\equiv\hat{\mathcal{U}}(N)\hat{S}_{1}(-r_1)\hat{S}_{2}(r_2)...\hat{S}_{N}(r_2)\ket{vac}_{1...N},
	\end{equation}                                                                                                                                     
	where $\hat{\mathcal{U}}(N)$ is the unitary as in \eqref{eq:psi} and can be created using the circuit in Fig. \ref{fig:wgarray} (see Appendix \ref{sec:def}). 
	
	The parameter $r_1$  (initial squeezing in mode $\hat{a}_1$), and $r_2$ (initial squeezing in all other modes) are significant from a practical perspective. Mode $\hat{a}_1$ could be squeezed off-chip, allowing a wider variety of techniques and larger pump powers than self-pumped modes. Thus, it is reasonable to expect architectures where the range of $r_1$ is much larger than $r_2$. As shown in the next two sections, there are cases where the gain is largest when $r_2$ is small compared to $r_1$, optimizing the enhancement of the circuit provided in Fig. \ref{fig:wgarray}. 
	
	\subsection{Entanglement in the initial state} \label{sec:Nmodes}
	\begin{figure*}[t!]
		\includegraphics[width=1.\textwidth]{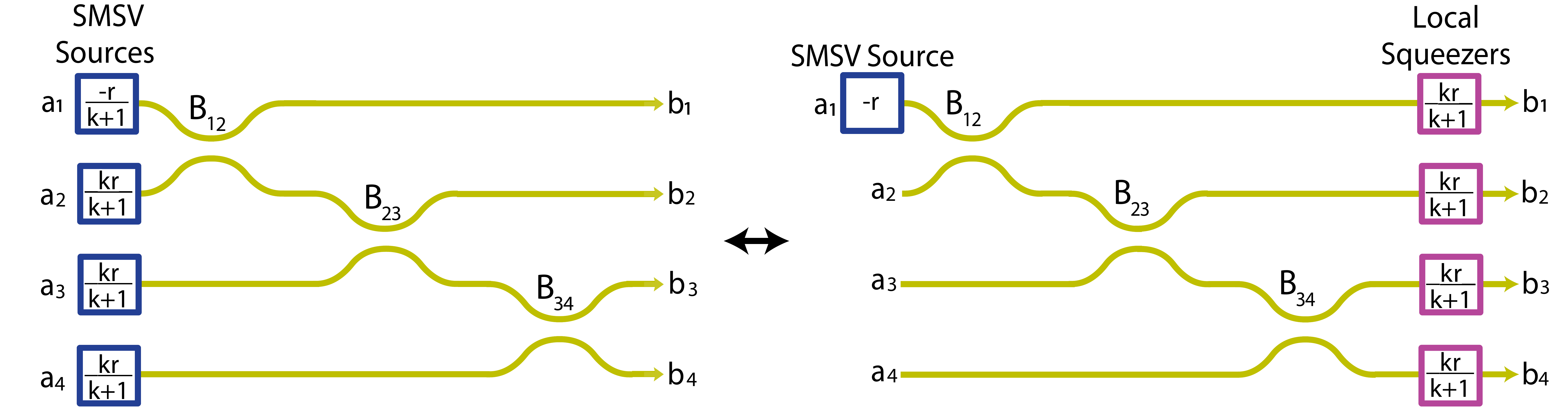}
		\caption{Circuit representation of Eq. \eqref{eq:localequiv}. The states $\ket{\phi_0\left(N,\frac{r}{k+1},\frac{kr}{k+1}\right)}$ (left) and $\ket{\psi_0(N,-r)}$  (right, before the local squeezers) are equivalent up to the local transformation $\ul$ represented by the local squeezers at the end of the circuit on the right.}
		\label{fig:equiv}
	\end{figure*}
	To allow the comparison of states with the same logarithmic negativity before photon subtraction, the state can be re-parametrized using parameters $r$ and $k$ such that $r_1=\frac{r}{k+1}$ and $r_2=\frac{kr}{k+1}$. Which, as seen in Appendix \ref{sec:def}, can be explicitly represented as 
	\begin{multline}\label{eq:phinorm}
		\ket{\phi_0\left(N, \frac{r}{k+1}, \frac{kr}{k+1}\right)} = \exp\left(\frac{r}{2}\left(\frac{\left(\frac{kN}{k+1}-1\right)\sum_{i=1}^N\hat{b}_i^2}{N}\right.\right.\\
		\left.\left.-\frac{2\sum_{i>j}\hat{b}_i\hat{b}_j}{N} \right) -h.c.\right) \ket{vac}_{1...N}.
	\end{multline}                                                                                                                                                                                                  
	In this notation, $k=0$ corresponds to a CV GHZ state prepared using a single source with squeezing $r$. The comparison is simplified due to the equivalence of states with different $k$ up to local unitaries 
	\begin{equation}\label{eq:localequiv}                                            
	\ket{\phi_0\left(N,\frac{r}{k+1},\frac{kr}{k+1}\right)} \lueq \ket{\psi_0(N,-r)},
	\end{equation}
	with $\ul=\left[\prod_{l=1}^N\hat{S}_{l}\left(\frac{kr}{k+1}\right)\right]$ 
	derived in Appendix \ref{sec:decomp} using the commutation relation $\left[x\sum_k(a_k)^2-(a_k^\dagger)^2\,,\,y\sum_{m>n}(a_ma_n)-(a^\dagger_ma^\dagger_n)\right]=0$ for all $x,y\in\mathbb{R}$. Eq. \eqref{eq:localequiv} can be mapped to an optical circuit equation depicted in Fig. \ref{fig:equiv}. This equality implies that the entanglement of the state $\ket{\phi_0\left(N,\frac{r}{k+1},\frac{kr}{k+1}\right)}$ is the same as the entanglement of the state $ \ket{\psi_0(N,-r)}$.
	
	\subsection{Entanglement in the photon subtracted state}
	The EvPS protocol is successful if a single photon is subtracted (i.e a single detector in Fig. \ref{fig:wgarray} fires). As before, the photon subtracted mode is labeled  $\hat{b}_A$ and 
	\begin{equation}
	\ket{\phi_1\left(N,\frac{r}{k+1},\frac{kr}{k+1}\right)}\equiv\hat{b}_A\ket{\phi_0\left(N,\frac{r}{k+1},\frac{kr}{k+1}\right)}.
	\end{equation}
	The partitions need to be defined in order to calculate bi-partite entanglement in this multimode state. The four composite modes ($A,B,C,D$) are  defined such that mode $A$ is the photon subtracted mode, $B$ are the modes coupled with $A$, while $C$ and $D$ are the groupings of modes needed for a bipartite splitting. Since the state $\ket{\phi_1\left(N,\frac{r}{k+1},\frac{kr}{k+1}\right)}$ is symmetric in all modes, all modes except $A$ will be equivalent and entanglement will depend only on the number of modes in each composite mode $B,C$ and $D$. In this way, the three generalized splittings that represent all possible splittings can be constructed as either: $(AB)_i-C_j$, $\mathrm{Tr}(D_k) (AB)_i-C_j$, or $\mathrm{Tr}((AB)_i) C_j-D_k$, where the subscripts represent the relative number of modes in each composite mode. For example, in the splitting $(AB)_{1/4}-C_{3/4}$ when $N=4$, $A$ is the single photon subtracted mode, $B$ does not contain any modes, and $C$ contains 3 modes, while if $N=100$ then $B$ would contain 24 modes and $C$ would contain 75 modes.
	
	Since entanglement is invariant under local unitary transformations, Eq. \eqref{eq:localequiv} can be used to see that entanglement in $\ket{\phi_1\left(N,\frac{r}{k+1},\frac{kr}{k+1}\right)}$ is equivalent to the entanglement in $\hat{S}_{A}^\dagger\left(\frac{kr}{k+1}\right)\hat{b}_A\hat{S}_{A}\left(\frac{kr}{k+1}\right)\ket{\psi_0(N,-r)}$ (and similarly a after partial trace). Using the relation $\hat{S_A}^{\dagger}(r)\hat{b}_A\hat{S}(r)=\hat{b}_A\cosh(r)-\hat{b}_A^{\dagger}e^{i\theta}\sinh(r)$ \cite{GandK}, both simplifies the calculation of logarithmic negativity in the photon subtracted  and provides intuition for the analysis of the performance of EvPS. The photon subtraction operation on the locally squeezed state has a different effect than the same operation on the state without local squeezing. 
	
	For any bi-partition, the logarithmic negativity of the state can be calculated using
	\begin{multline}\label{eq:phi3mode}
		\left[\hat{b}_A \cosh(\frac{kr}{k+1})-\hat{b}_A^\dagger \sinh(\frac{kr}{k+1})\right]\\
		\times\exp\left(\frac{-r}{2}\left(\frac{\hat{b}_A+\sqrt{n-1}\hat{b}_{B}}{\sqrt{N}}\right.\right.\\
		\left.\left.+\frac{\sqrt{m}\hat{b}_{C}}{\sqrt{N}}\right)^2-h.c.\right)\ket{vac}_{A,B,C},
	\end{multline}
	where $\hat{b}_{B}=\frac{\hat{b}_{i_{1}}+...+\hat{b}_{i_{n}}}{\sqrt{n}}$ and $\hat{b}_{C}=\frac{\hat{b}_{j_{1}}+...+\hat{b}_{j_{m}}}{\sqrt{m}}$, where $n+m+1=N$ and $i_1,...,i_n$ and $j_1,...,j_m$ represents all the possible orderings of the $N-1$ modes left for a given mode $A$. An extension to include traced-out mode is included in Appendix \ref{sec:traces}.
	
	\section{Performance of the scheme} \label{sec:performance}
	Using the tools above, an exploration of the parameter space and a locating of points and regions where the advantage is optimal regarding both the input squeezing parameters and the number of modes can be pursued. While the symmetric nature of the states allows a significant simplification of the analytical expressions, the comparisons were made numerically in Python using QuTiP \cite{qutip}. This numerical method requires a cutoff at high photon numbers and consequently introduces larger errors at higher squeezing parameters. Hence, the squeezing parameter, $r$, was chosen to have a negligible numerical error while upholding a minimal detection probability. However, the range of the squeezing parameter ratio, $k=r_2/r_1$, and the number of modes, $N$, are chosen to be exemplative of the full range of behaviors. The results presented in the main text concern pure states splittings;  results for mixed states splitting (by tracing out some modes) are presented in Appendix \ref{sec:traces}. 
	
	\subsection{Optimizing the parameters}\label{subsec:opt}
	\begin{figure}
		\includegraphics[width=\columnwidth]{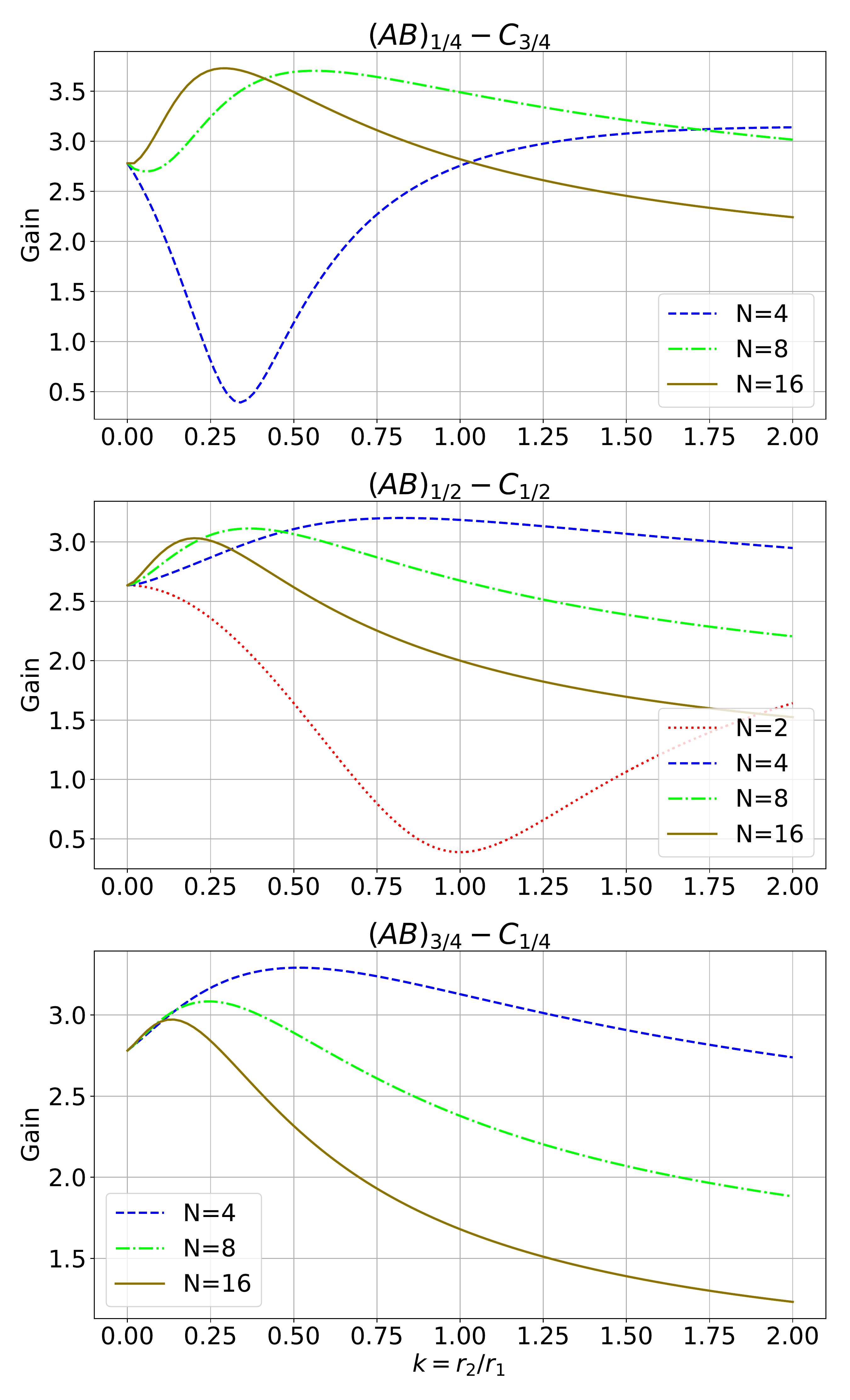}\caption{The gain in entanglement for a $N$-mode state with squeezing parameter $r=0.2$, $\ket{\phi_1\left(N,\frac{0.2}{k+1},\frac{k(0.2)}{k+1}\right)}$, is plotted against $k$, the ratio between the input squeezing parameter in mode $\hat{a}_1$ and squeezing parameter in the $N-1$ other modes. These single mode squeezed states are passed through directional couplers to generate CV GHZ states, $\ket{\phi_0\left(N,\frac{0.2}{k+1},\frac{k(0.2)}{k+1}\right)}$, which sees an enhancement in the entanglement after photon subtraction. In this figure, only the splitting of the form $(AB)_i-C_j$ are considered, where mode $A$ is assumed to be the photon subtracted mode, while $B$ and $C$ are the groupings of modes such that $i$ and $j$ in $(AB)_{i}$ and $C_j$ represent the ratio of modes contained in the given splitting. Bipartite splitting containing traced out modes can been seen in Fig. \ref{fig:k-N_opt-2} in Appendix \ref{sec:traces}.}
		\label{fig:k-N_opt}
	\end{figure}
	To show how the additional sources influence the EvPS procedure, the gain was studied as a function of $k$ for various values of $N$. Results for $N=2,4,8,16$ at $r=0.2$ are plotted in Fig. \ref{fig:k-N_opt} (Fig. \ref{fig:k-N_opt-2} for the partial traced states). As can be expected in the limit of $k=0$, the gain is independent of $N$. On the other hand, when $k\gg1$ an asymptotic behavior is expected to arise as $\ket{\phi_1(4, \frac{r}{k+1}, \frac{kr}{k+1})} \rightarrow \ket{\phi_1(4, 0, r)}$. 
	
	Furthermore, in Fig. \ref{fig:k-N_opt} a significant dip in the logarithmic negativity occurs for $N=2$ and $N=4$ in the $(AB)_{1/2}-C_{1/2}$ and $(AB)_{1/4}-C_{3/4}$ splittings respectively (i.e., when $AB$ is a single mode). These dips are due to the single mode terms, $\left(\frac{kN}{k+1}-1\right)\sum_{i=1}^{N}\hat{b}_i$, in Eq. \eqref{eq:phinorm} decaying to zero when $k=\frac{1}{N-1}$. These single mode terms affect the subsequent photon subtraction operation and can increase gain \cite{LocalSq}. The dip is counteracted when other modes are grouped with the photon subtracted mode, hence the lack of a dip in the gain in the $(AB)_{3/4}-C_{1/4}$ splitting. An important consequence of this result is that as the number of modes is increased the smaller $\frac{1}{N-1}$ gets; therefore, the dip moves to smaller $k$ as the number of modes increases, enabling higher levels entanglement to be reached for smaller values of $k$. 
	
	Surprisingly, in the two-modes case, the gain is optimal when $k=0$ (i.e., a single source). Moreover, for four parties, at any $k>0$ there is at least one bi-partition such that the gain in entanglement will be smaller than the $k=0$ case. However, this is not generic since for $4n$-modes, $\{n>1, n\in\mathbb{Z}\}$, there are values of $k>0$ such that the gain in entanglement surpasses the one of $k=0$ for all bipartite splitting of a four-party scheme, demonstrating a definite increase in the multipartite entanglement. 
	
	Since the gain depends on the partition, the optimal value of $k$ depends on the particular figure of merit. One optimal value is the maximal gain for any bi-partition (in this case at $r=0.2$)  given at $N=4,\; k\approx0.82$ for $(AB)_{1/2}-C_{1/2}$. A second is the optimal gain region, corresponding the parameter space where the gain is larger than that of a single source for all bi-partitions of a four-party scheme. For these results, all possible bi-partitions involving at least $1/4$ of the modes were considered, and for $r=0.2$, and $N=8$ the optimal region of $0.21\le k\le0.33$ was found, for our purposes $k=0.24$ was chosen as an explicit example in further plots. 
	
\begin{figure}[h!]
		\includegraphics[width=\columnwidth]{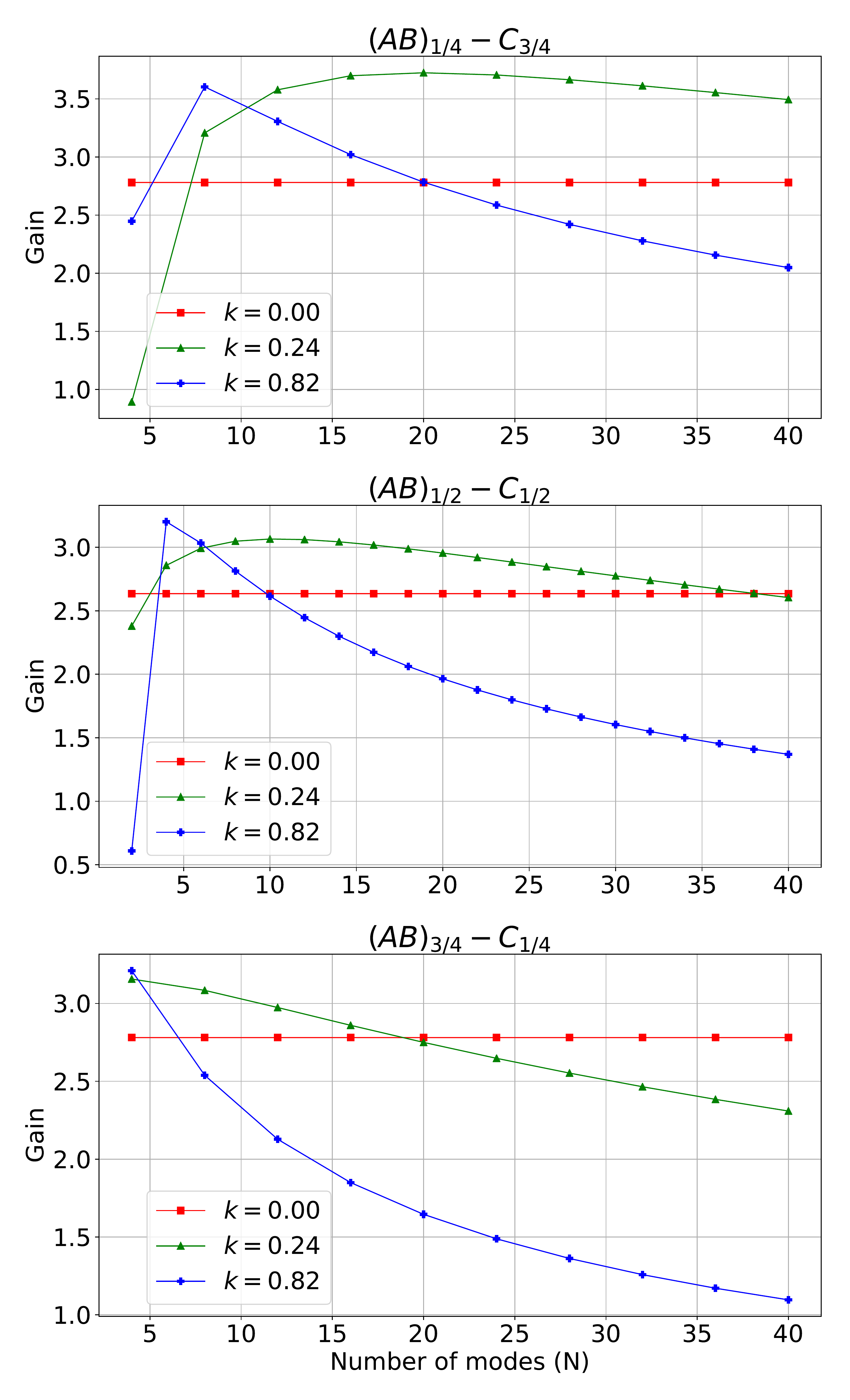}
		\caption{The gain in entanglement of the photon subtracted state, $\ket{\phi_1\left(N, \frac{r}{k+1}, \frac{kr}{k+1}\right)}$ is plotted as a function of the number of modes $N$ for values of $k=0$, $0.24$ and $0.82$ (squares, triangles and diamonds respectively, solid lines are only a guide to the eye). In each plot, mode $A$ is the photon subtracted mode, while $B$ and $C$ are the groupings of modes such that $i$ and $j$ in $(AB)_{i}-C_j$ represent the ratio of modes contained in the given splitting. Note that in plots $(AB)_{1/4}-C_{3/4}$ and $(AB)_{3/4}-C_{1/4}$, the number of modes, $N=4n$, while for the plot $(AB)_{1/2}-C_{1/2}$, $N=2n$, where $n\in\mathbb{N}$. All result are evaluated at a squeezing parameter value of $r=0.2$. In this figure, only the pure state splittings are reviewed, for the full set of splittings Fig. \ref{fig:Nopt-2} in Appendix \ref{sec:traces} can be consulted.}
		\label{fig:Nopt}
	\end{figure}

	By varying the number of modes, it becomes apparent that the optimal value does not have to be very large and that the gain is not monotonic in the number of modes. The gain as a function of the number of modes is plotted in Fig. \ref{fig:Nopt} (Fig. \ref{fig:Nopt-2} for the partially traced state) for $r=0.2$ and three values of $k=0,0.24,0.82$ corresponding to the single source state and the two optimal values above. 
	
	\subsection{Losses}\label{sec:loss}                                    
\begin{figure}[t!]
		\includegraphics[width=\columnwidth]{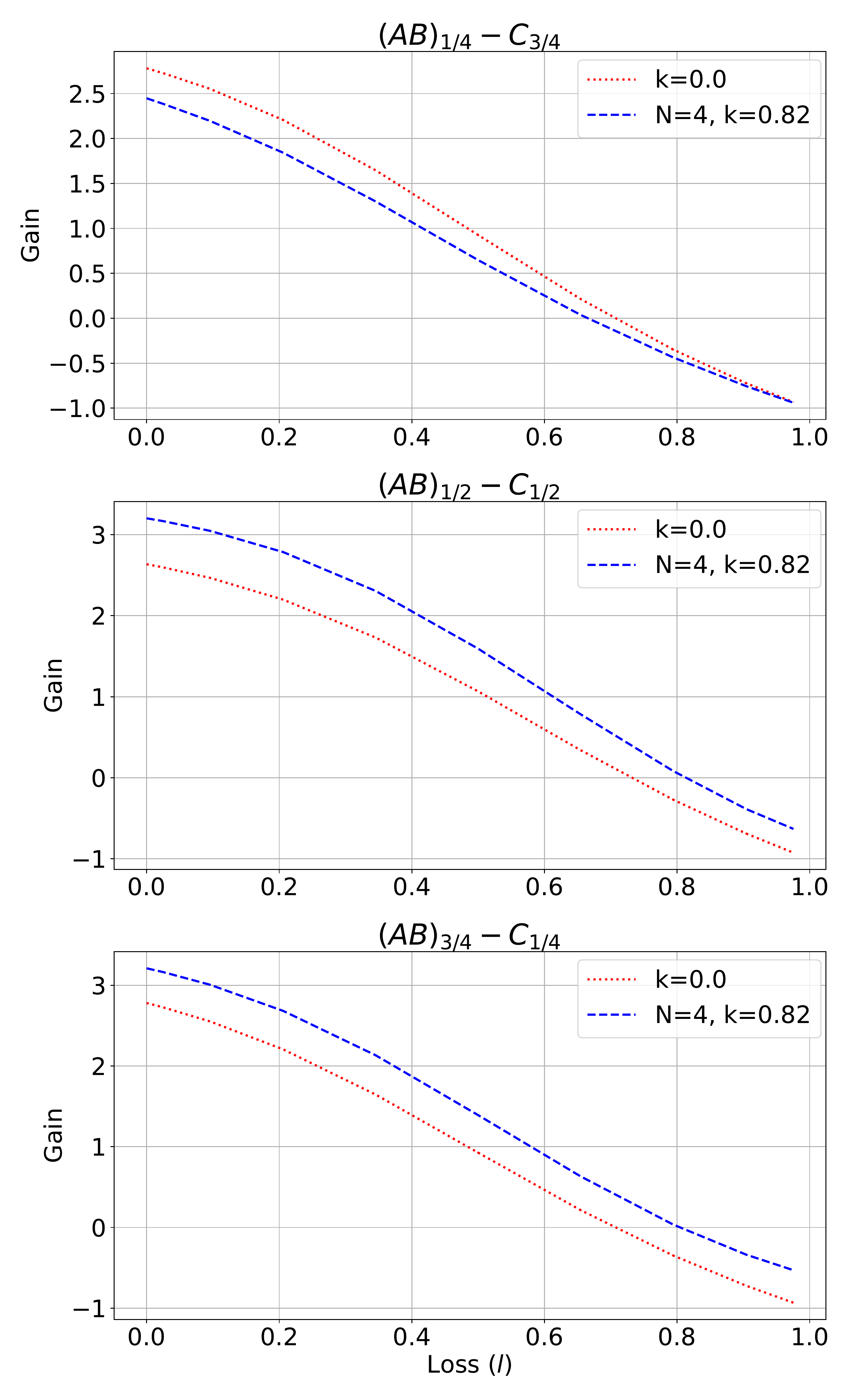}
		\caption{The effects of loss on the gain in entanglement of the four mode CV GHZ state with squeezing parameter $r=0.2$ and ratios between squeezing parameters, $k=0.0, 0.82$, $\ket{\phi_1\left(4,\frac{0.2}{k+1},\frac{k(0.2)}{k+1}\right)}$ are studied in this figure. The loss is considered to have been applied after the generation of the CV GHZ state, but before the photon subtraction, as shown in Fig. \ref{fig:wgarray}. In this figure, the mode $A$ is the photon subtracted mode, and $B$ and $C$ are the groupings of modes such that $i$ and $j$ in $(AB)_{i}$ and $C_j$ represent the ratio of modes contained in a splitting. Note that for the $k=0$ curve (red), $N$ is not specified since the single source case is independent of $N$.}
		\label{fig:losses}
	\end{figure}

	In subsection \ref{subsec:opt}, EvPS was studied under ideal conditions under the assumption that the main experimental constraint would be in generating higher levels of squeezing. Realistically, the loss would be another major factor in limiting entanglement of the outgoing state. The effect of loss is considered on the CV GHZ state before the photon subtraction operation, as seen in Fig. \ref{fig:wgarray}, under the assumption that loss is equivalent for all modes. The loss in each mode was simulated using a BS transformation and an ancillary mode so that the state after the loss could be explicitly written as
	\begin{multline}
		\rho_{out} = \mathrm{Tr}_{1'\,2'\,3'\,4'}\left(\BS{1}{1'}{\theta}\BS{2}{2'}{\theta}\BS{3}{3'}{\theta}\BS{4}{4'}{\theta}\right.\\
		\left. \ketbra{\phi_0\left(4,\frac{0.2}{k+1},\frac{(0.2)k}{k+1}\right)}{\phi_0\left(4,\frac{0.2}{k+1},\frac{(0.2)k}{k=1}\right)} \right.\\\
		\left.\BSd{1}{1'}{\theta}\BSd{2}{2'}{\theta} \BSd{3}{3'}{\theta} \BSd{4}{4'}{\theta} \right),
	\end{multline}
	where $k=0$ or $k=0.82$ as seen in Fig. \ref{fig:losses}, modes $1',2',3',$ and $4'$ are the modes in which the losses are coupled, and the loss parameter is given by $l=\sin^2(\theta), \theta \in [0, \frac{\pi}{2}]$.
	
	As before, the photon subtracted state is $a_A\rho_{out}a_A^\dagger$ and gain was calculated with respect to $\rho_{out}$.  The results show that in this regime, the multiple source ($k>0$) states can outperform the single source ($k=0$) states at the same settings as in the lossless case. The gain after loss for $N=4$ with a squeezing parameter of $r=0.2$ and the optimal value $k=0.82$ (see Fig. \ref{fig:k-N_opt}) and $k=0$ are plotted in Fig. \ref{fig:losses} (Fig. \ref{fig:lossesas} for the partial traced state). The gain is positive in the $(AB)_{1/2}-C_{1/2}$  (for the above values) for a loss parameter of up to $l=0.81$, but a positive gain in all splittings requires a loss parameter of $l\le0.36$. Above that threshold, photon subtraction has the opposite effect, i.e., reducing entanglement. These values can be compared with the single source case where corresponding thresholds are $l=0.73$ and $l\le0.46$ respectively.
	
	\section{Discussion and conclusion} \label{sec:Discussion}
	Motivated by recent developments in the generation of multi-mode squeezed states on-chip, the performance of the EvPS protocol was studied and compared on states generated by photonic circuits with single and multiple squeezed inputs. The objective was to give preliminary answers to questions related to practical experimental challenges. In particular, it was of interest to find whether there is an advantage in using multiple sources and if this advantage could be significant even in the case where there are a few modes and/or when the additional squeezed states have a smaller squeezing parameter. These results show that, at least for the subsets of states under consideration, the answer is yes; that is, there is a significant advantage even when there are a few additional modes with less squeezing at the inputs. Moreover, more is usually not better, i.e., there is a finite number of modes that is optimal, and the optimal squeezing ratio, $k$, is usually smaller than one. 
	
	The analysis was restricted to a subset of CV GHZ states which could be generated by the circuit in Fig. \ref{fig:wgarray}. While this design was chosen for its experimental feasibility, it could be modified to generate any Gaussian state \cite{Aharon}; however, the number of free parameters and the computational difficulty would make the analysis for all Gaussian states difficult. The particular three-parameter family studied for the state, $\ket{\phi_0(N,\frac{r}{k+1},\frac{kr}{k+1})}$, is relatively simple to analyze due to the equivalence discussed in Sec. \ref{sec:Nmodes} (see also Fig. \ref{fig:equiv}). Moreover, the free parameters represent the three main experimental challenges: extending the number of modes $N$, increasing squeezing in the externally pumped mode $r/(k+1)$, and increasing the (relative) squeezing, $k$, in the self-pumped modes. The results are promising on all fronts.
	
	Entanglement in the initial state is independent of $k$, so the relation between the squeezing values in mode $\hat{a}_1$ and the other modes implies a trade-off between squeezing in the externally pumped source and the self-pumped sources. For example, for a fixed value of entanglement, it is possible to increase one and decrease the other. However, there is an optimal value of the ratio $k$ for EvPS. Somewhat surprisingly, this value is not $k=1$ (see for example Fig. \ref{fig:k-N_opt}) and depends on the particular bi-partition which will ultimately depend on how the modes are distributed between parties. The fact that the optimal value for $k$ could be small (see for example $N=16$ in Fig. \ref{fig:k-N_opt}) means that an advantage can be observed when the self-pumped sources are significantly less powerful than the external source.
	
	The advantage of the multiple source setup can be examined through the circuit equality in Fig. \ref{fig:equiv}. The additional squeezed vacuum sources can be mapped to squeezers at the end of the circuit. Although these local squeezers do not increase the entanglement, they do have an effect on the subsequent subtraction by inducing the mapping $\hat{b}_A\rightarrow \hat{b}_A \cosh(\frac{kr}{k+1})-\hat{b}_A^\dagger \sinh(\frac{kr}{k+1})$. However, this mapping should be applied in the same way to all operations following preparation and in particular to loss between the sender and the receiver. As shown in Sec. \ref{sec:loss} (See Fig. \ref{fig:losses} and \ref{fig:lossesas}) the loss does not have a significantly worse effect on EvPS when the $k>0$. 
	
	One somewhat unexpected result is that the performance does not increase monotonically with the number of modes, and in-fact for larger values of $k$ it peaks at relatively low $N$ (see Fig. \ref{fig:Nopt}). Hence, in practice, it would be possible to experimentally observe the advantages of the multiple source protocol using relatively small devices. The technology to implement the building blocks for such a device has already been demonstrated (see \cite{Aharon} for review of recent results) and demonstrations of such integrated devices should appear shortly. Such a demonstration would provide further motivation to study the advantages of using multiple squeezing sources, beyond the EvPS protocol. 
	
	\begin{acknowledgments}
		We are grateful to W. Wu for his useful comments on this manuscript. 
	\end{acknowledgments}
	
	\appendix
	\section{CV GHZ states}\label{sec:def}
	\begin{figure}[h!]
		\includegraphics[width=\columnwidth]{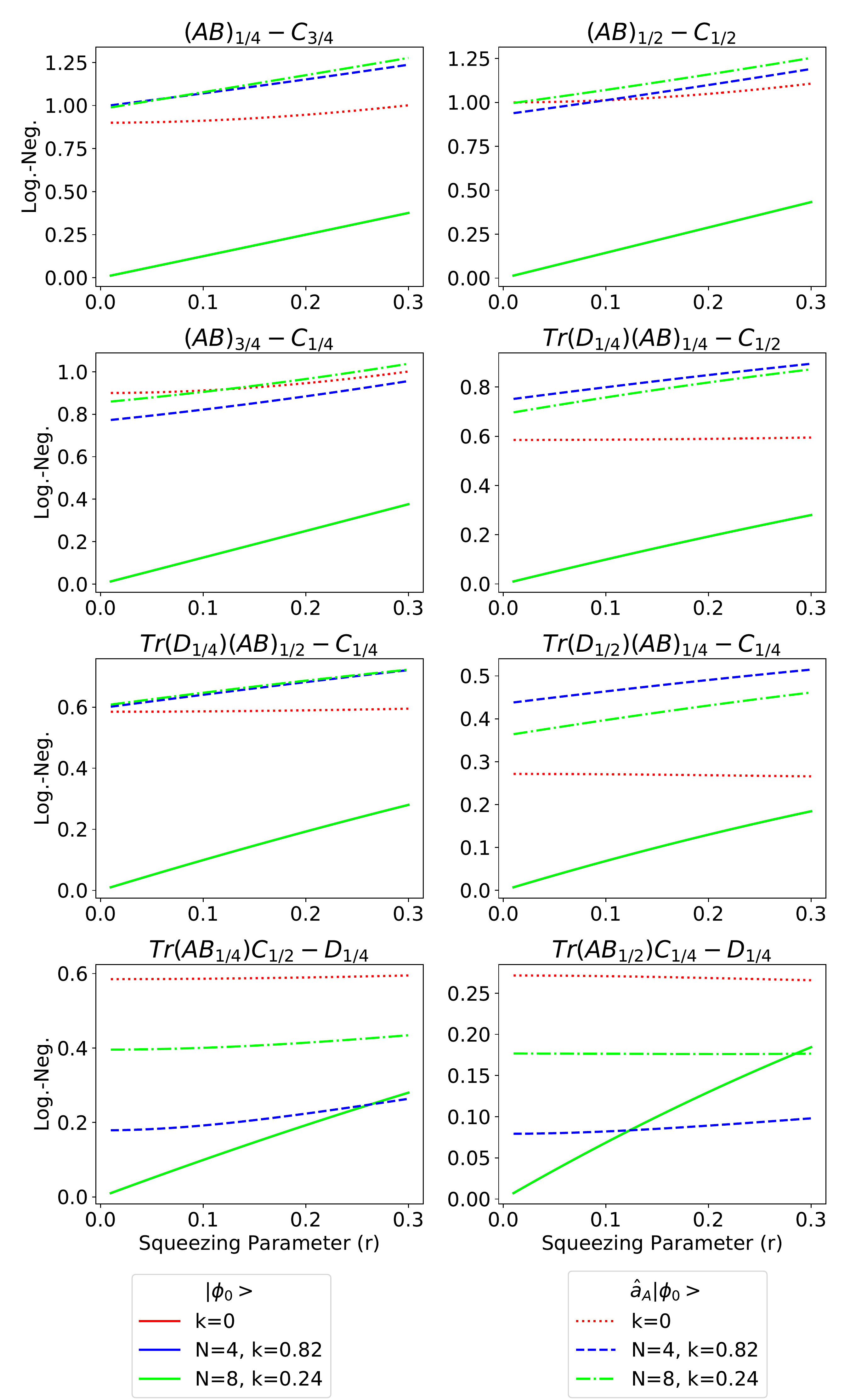}
		\caption{The logarithmic negativity of the single photon subtracted CV GHZ state, $\hat{a}_A\ket{\phi_0\left(N,\frac{r}{k+1},\frac{kr}{k+1}\right)}\equiv\ket{\phi_1(N,\frac{r}{k+1},\frac{kr}{k+1})}$, is contrasted to that of the CV GHZ state before photon subtraction, $\ket{\phi_0\left(N,\frac{r}{k+1},\frac{kr}{k+1}\right)}$, when $k$ takes on the optimums found in Fig. \ref{fig:k-N_opt} for when $N=2,4,$ and $8$, and is plotted against the squeezing parameter, $r$. In this figure, the logarithmic negativity of all possible splittings are considered, and mode $A$ is assumed to be the photon subtracted mode, while $B$ and $C$ are the groupings of modes such that $i$ and $j$ in $(AB)_{i}$ and $C_j$ represent the ratio of modes contained in the given splitting. Note that for the $k=0$ curve (red), $N$ is not specified since the single source case is independent of N.}
		\label{fig:E_Nvr}
	\end{figure}
	The $N$ mode symmetric splitter consisting of $N-1$ directional couplers, as depicted in Fig. \ref{fig:wgarray}, is simpler to analyze using the Bogoliubov transformation
	\begin{equation}
	\begin{pmatrix}
	\hat{b}_1 & ... & \hat{b}_N
	\end{pmatrix}^T
	=\tilde{\mathcal{U}}(N)
	\begin{pmatrix}
	\hat{a}_1 & ... & \hat{a}_N
	\end{pmatrix}^T,
	\end{equation}
	where $\tilde{\mathcal{U}}(N)$ is the unitary transformation corresponding to the series of beam splitter operations
	\begin{equation}\label{eq:Utrans}
	\tilde{\mathcal{U}}(N)=\tilde{B}_{N-1\,N}\left(\arcsin\frac{1}{\sqrt{2}}\right)...\tilde{B}_{12}\left(\arcsin\frac{1}{\sqrt{N}}\right).
	\end{equation}
	To relate the interferometer transformation \eqref{eq:Utrans} to the integrated setting the general description of a directional coupler can be found in \cite{YarivAndYeh}, and its dispersion properties allowing for quantum state engineering applications unavailable in bulk optics can be found in \cite{RyanLPR, RyanOptica}. Here the phase-free directional coupler operation on two modes with respective annihilation operators $\hat{a}_i$ and $\hat{a}_j$ is considered which follow the notation convention from \cite{QIwCV} 
	\begin{equation}\label{eq:BS}
	\begin{pmatrix}
	\hat{b}_i \\ \hat{b}_j
	\end{pmatrix}
	=
	\begin{pmatrix}
	\sin\theta & \cos\theta \\
	\cos\theta & -\sin\theta
	\end{pmatrix}
	\begin{pmatrix}
	\hat{a}_i \\ \hat{a}_j
	\end{pmatrix}.
	\end{equation}
	The matrix $\tilde{B}_{ij}(\theta)$ representing the mode transformation given by the directional coupler is given by the identity matrix with the elements $\mathbb{I}_{ii}$, $\mathbb{I}_{ij}$, $\mathbb{I}_{ji}$, and $\mathbb{I}_{jj}$ are replaced by the corresponding entries of Eq. \eqref{eq:BS}. Alternatively, this transformation may be written in the Heisenberg picture formulation of the beam splitter which gives the transformation
	\begin{equation}
	\begin{pmatrix}
	\hat{b}_i \\ \hat{b}_j
	\end{pmatrix}
	=
	\hat{B}(\theta)^\dagger
	\begin{pmatrix}
	\hat{a}_i \\ \hat{a}_j
	\end{pmatrix}
	\hat{B}(\theta),
	\end{equation}
	where $\hat{B}(\theta)$ is the unitary given by 
	\begin{equation}
	\hat{B}(\theta)=\exp\left[\left(\theta-\frac{\pi}{2}\right)\left(\hat{a}_0\hat{a}_1^\dagger-\hat{a}_0^\dagger\hat{a}_1\right)\right].
	\end{equation}
	
	Writing the input modes in terms of the output modes gives 
	\begin{equation}\label{eq:bogo}
	\begin{pmatrix}
	\hat{a}_1 \\ \hat{a}_2 \\ \hat{a}_3 \\ ... \\ \hat{a}_{N-1} \\ \hat{a}_N
	\end{pmatrix}
	=\tilde{\mathcal{U}}^{-1}(N)
	\begin{pmatrix}
	\hat{b}_1 \\ \hat{b}_2 \\ \hat{b}_3 \\ ... \\ \hat{b}_{N-1} \\ \hat{b}_N
	\end{pmatrix}
	=\begin{pmatrix}
	\frac{\hat{b}_1+ ... +\hat{b}_N}{\sqrt{N}} \\
	\frac{(N-1)\hat{b}_1-\hat{b}_2- ... -\hat{b}_N}{\sqrt{(N-1)^2+(N-1)}} \\
	\frac{(N-2)\hat{b}_2-\hat{b}_3- ... -\hat{b}_N}{\sqrt{(N-2)^2+(N-2)}} \\
	...\\
	\frac{2\hat{b}_{N-2}-\hat{b}_{N-1}-\hat{b}_N}{\sqrt{6}}\\
	\frac{\hat{b}_{N-1}-\hat{b}_N}{\sqrt{2}}
	\end{pmatrix}.
	\end{equation}
	
	Note that the relation between the Heisenberg picture operator $\hat{\mathcal{U}}(N)$ and the Bogoliubov transformation $\tilde{\mathcal{U}}(N)$ can be obtained through 
	\begin{equation}
	\begin{pmatrix}
	\hat{b}_1 & ... & \hat{b}_N
	\end{pmatrix}^T
	=\hat{\mathcal{U}}^{\dagger}(N)
	\begin{pmatrix}
	\hat{a}_1 & ... & \hat{a}_N
	\end{pmatrix}^T
	\hat{\mathcal{U}}(N)
	\end{equation}
	and one can replace the $\tilde{U}$ and $\tilde{B}$'s with $\hat{U}$ and $\hat{B}$'s in \eqref{eq:Utrans} to obtain the explicit formulation in this picture.
	
	\subsection{The unitary equivalence between $\ket{\phi_0}$ and $\ket{\psi_0}$}\label{sec:decomp}
	To derive \eqref{eq:localequiv} in the main text, the final state can be written in terms of the output $\hat{b}_i$ and $\hat{b}_i^\dagger$ operators and through the use of the Baker-Hausdorf-Campbell theorem the local squeezing terms can be isolated. This is possible through the commutation  relation 
	\begin{multline}\label{eq:BHC}
		\left[\left( \sum_{i=1}^{N}x\hat{b}_i^2\right) -h.c., \left( \sum_{i>j}^{N}y\hat{b}_i\hat{b}_j\right) -h.c. \right]\\
		=2\sum_{i=1}^{N}x\hat{b}_i\left( \sum_{\overset{j=1}{j\neq i}}^N y\hat{b}_j^\dagger\right) - 2\sum_{i=1}^{N}x\hat{b}_i^\dagger\left( \sum_{\overset{j=1}{j\neq i}}^N y\hat{b}_j\right)=0,
	\end{multline}
	for $x,y\in\mathbb{R}$. 
	
	Applying the result of Eq. \eqref{eq:bogo} to transform \eqref{eq:GHZ} and inserting $r_1=\frac{r}{k+1}, r_2=\frac{kr}{k+1}$ gives 
	\begin{align}
		\ket{\phi_0\left(N,\frac{r}{k+1},\frac{kr}{k+1}\right)} =e^{V-V^\dagger}\ket{vac}
	\end{align}
	with 
	\begin{multline}\label{eq:vfact}
		V=\frac{-r}{2(k+1)}\frac{\left(\sum_{k=1}^N\hat{b}_k\right)^2}{N}\\
		+\frac{kr}{2(k+1)}\sum_{l=1}^{N-1}\frac{\left[(N-l)\hat{b}_{l}-\sum_{j=l+1}^{N}\hat{b}_j\right]^2}{(N-l)^2+N-l},
	\end{multline}
	which can be simplified using
	\begin{multline}
		\sum_{l=1}^{N-1}\frac{\left[(N-l)\hat{b}_l-\sum_{j=l+1}^{N}\hat{b}_j\right]^2}{(N-l)^2+N-l}\\
		=\left(1-{1\over N}\right)\sum_{l=1}^{N}\hat{b}_l^2 -\frac{2}{N}\sum_{l=1}^{N}\sum_{j>l}\hat{b}_j\hat{b}_l.
	\end{multline}
	Inserting the above back into \eqref{eq:vfact} isolates the terms that are independent of $k$ which commute with the rest of the $k$ (see Eq. \eqref{eq:BHC}) and the state can be brought into the final form,
	\begin{align}
		&\ket{\phi_0 \left(N,\frac{r}{k+1},\frac{kr}{k+1}\right)} \nonumber \\
		& =\left[\prod_{l=1}^N\hat S_l\left(\frac{kr}{k+1}\right)\right]\hat{\mathcal{U}}(N)\hat S_1(-r)\ket{vac} \nonumber\\
		& =\left[\prod_{l=1}^N\hat S_l\left(\frac{kr}{k+1}\right)\right] \ket{\psi_0(N,-r)}.
	\end{align}
	Allowing for the simplification of the photon subtracted state
	\begin{multline}\label{eq:ph-sub}
		\ket{\phi_1 \left(N,\frac{r}{k+1},\frac{kr}{k+1}\right)} \\
		= \hat{b}_A\left[\prod_{l=1}^N\hat S_l\left(\frac{kr}{k+1}\right)\right] \ket{\psi_0(N,-r)}\\
		=\prod_{l=1}^N\hat{S}_l\left(\frac{kr}{k+1}\right)\left((\hat{b}_A\cosh\left(\frac{kr}{k+1}\right)\right.\\
		\left.-\hat{b}_A^\dagger\sinh\left(\frac{kr}{k+1}\right)\right)\ket{\psi_0\left(N,-r\right)},
	\end{multline}
	where $A$ is the photon subtracted mode.
	
	The equivalence in logarithmic negativity can be seen in Fig. \ref{fig:E_Nvr} by noting that the curves plotting logarithmic negativity before photon subtraction (solid lines) are all overlapping; however, the photon subtraction operation exhibits different behaviors depending on the value of $k$ and $N$.
	
	It is also worth noting that as $r\rightarrow0$ the logarithmic negativity of the state before photon subtractions tends to zero, while the state after photon subtraction tends to values larger than zero. This is understood with our analysis in Eq. \eqref{eq:Bell}, while noting that although the gain would be large, the probability of a successful detection would be too low to be practical. 
	\begin{figure}[t!]
		\includegraphics[width=\columnwidth]{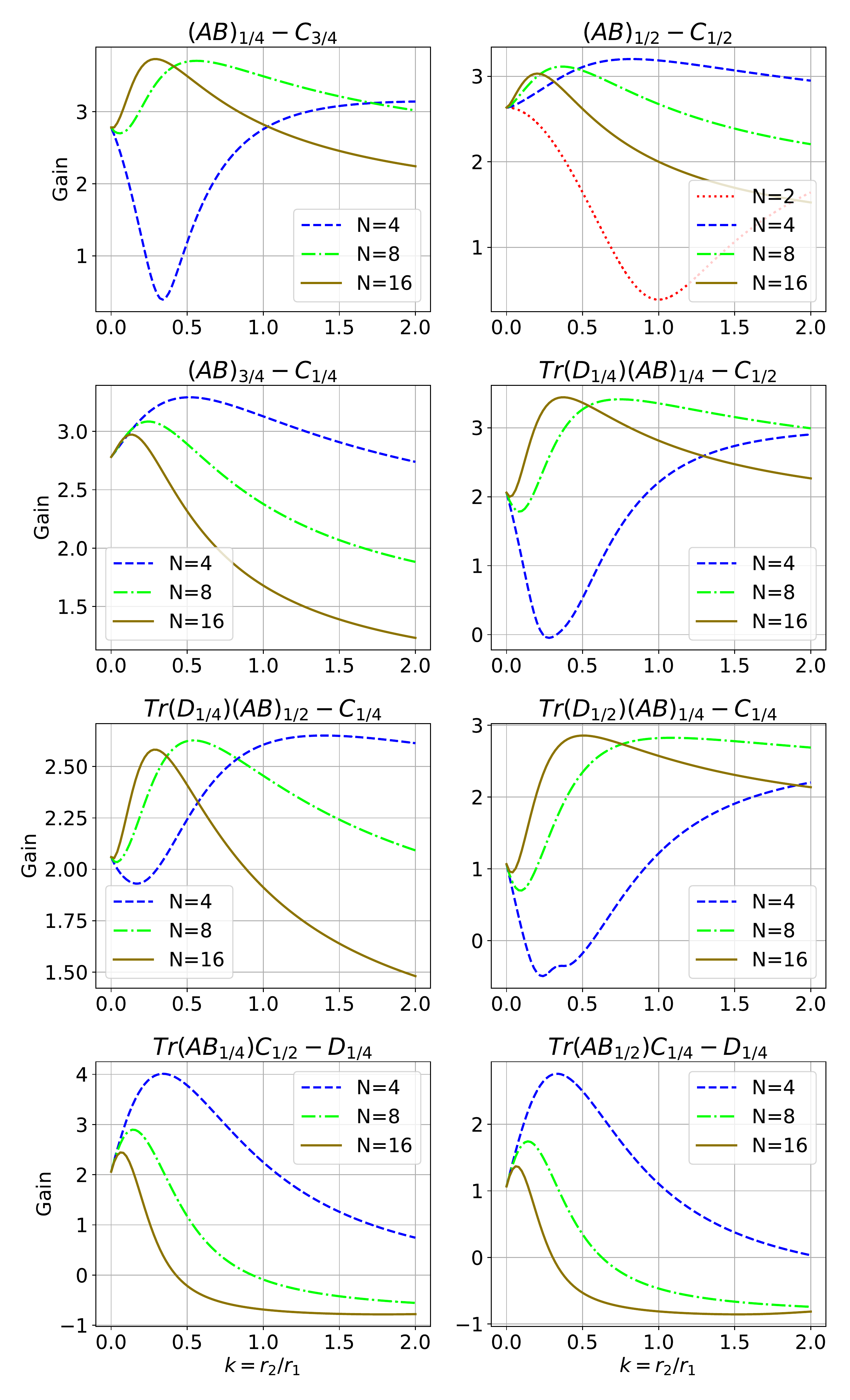}
		\caption{This figure is an extension of  Fig.  \ref{fig:k-N_opt} for all bipartite splittings.The gain in entanglement for the $N$-mode state with squeezing parameter $r=0.2$, $\ket{\phi_1\left(N,\frac{0.2}{k+1},\frac{k(0.2)}{k+1}\right)}$, is  plotted against, $k$, the ratio between the input squeezing parameter in mode $\hat{a}_1$ and the squeezing parameter from the other $N-1$ single mode squeezed vacuum (SMSV) inputs. These SMSV states are passed through directional couplers used to generate a CV GHZ state, $\ket{\phi_0\left(N,\frac{0.2}{k+1},\frac{k(0.2)}{k+1}\right)}$, which sees an enhancement in the entanglement after photon subtraction. In this figure mode $A$ is assumed to be the photon subtracted mode, while $B$ and $C$ are the groupings of modes such that $i$ and $j$ in $(AB)_{i}$ and $C_j$ represent the ratio of modes contained in the given splitting. }
		\label{fig:k-N_opt-2}
	\end{figure}
	
	\begin{figure}[t!]
		\includegraphics[width=\columnwidth]{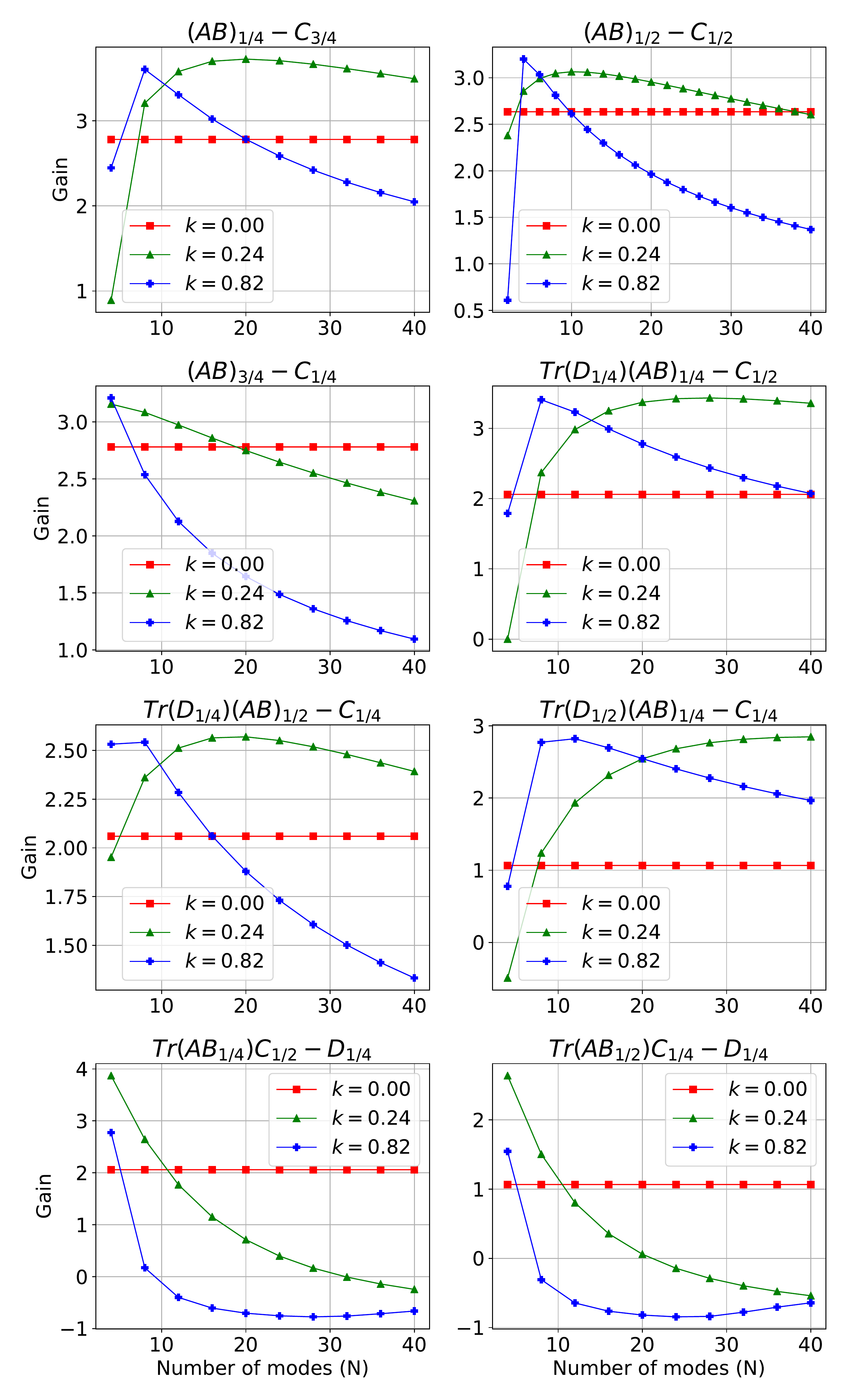}
		\caption{This figure is an extension of  Fig. \ref{fig:Nopt} for all bipartite splittings. The gain in entanglement of the photon subtracted state, $\ket{\phi_1\left(N, \frac{r}{k+1}, \frac{kr}{k+1}\right)}$ is plotted as a function of the number of modes $N$ for values of $k=0$, $0.24$ and $0.82$ (squares, triangles and diamonds respectively, solid lines are only a guide to the eye). In each plot, mode $A$ is the photon subtracted mode, while $B$ and $C$ are the groupings of modes such that $i$ and $j$ in $(AB)_{i}-C_j$ represent the ratio of modes contained in the respective splitting. Note that for the plot $(AB)_{1/2}-C_{1/2}$, the number of modes, $N=2n$ while in all other plots, $N=4n$, where $n\in\mathbb{N}$. All result are evaluated at a squeezing parameter value of $r=0.2$.}
		\label{fig:Nopt-2}
	\end{figure}
	
	\begin{figure}[h!]
		\includegraphics[width=\columnwidth]{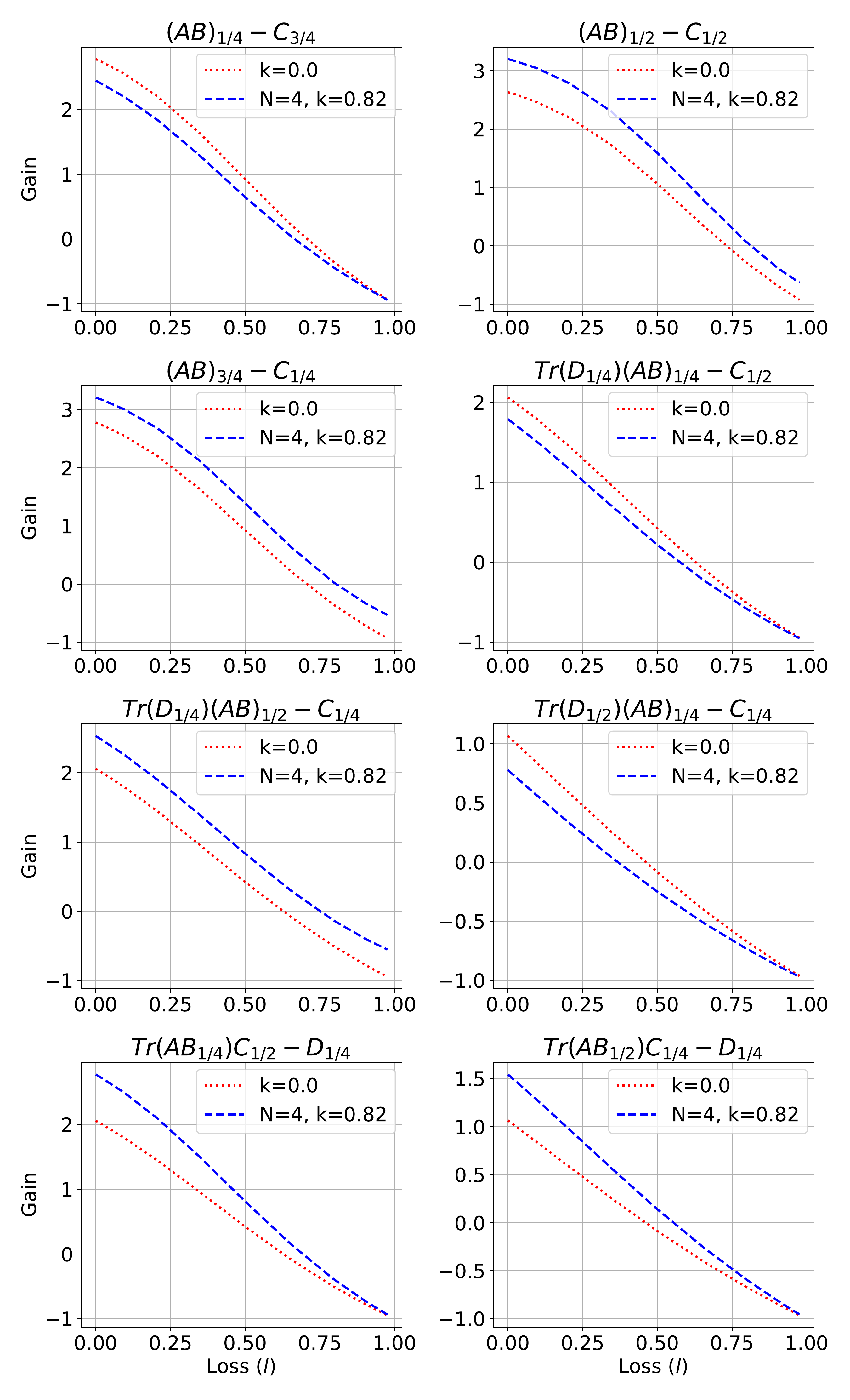}
		\caption{This figure is an extension of  Fig. \ref{fig:losses} for all bipartite splittings. The effects of loss on the gain in entanglement of the four-mode CV GHZ state with squeezing parameter $r=0.2$ and the ratios between squeezing parameters, $k=0, 0.82$, $\ket{\phi_1\left(4,\frac{0.2}{k+1},\frac{(0.2)k}{k+1}\right)}$. The loss is considered to have been applied after the generation of the CV GHZ state, but before the photon subtraction, as shown in Fig. \ref{fig:wgarray}. In this figure, the mode $A$ is the photon subtracted mode, and $B$ and $C$ are the groupings of modes such that $i$ and $j$ in $(AB)_{i}$ and $C_j$ represent the ratio of modes contained in a splitting.  Note that for the $k=0.0$ curve (red), $N$ is not specified since there is only a single source, hence it is independent of $N$.}
		\label{fig:lossesas}
	\end{figure}
	
	\section{Logarithmic negativity}\label{sec:Log-Neg}
	The logarithmic negativity  is defined as the base two logarithm of the sum of the absolute value of the eigenvalues of $\rho^{T_A}$, where $\rho^{T_A}$ is the partial transpose with respect to a subsystem $A$ of the density matrix, $\rho$. Using the trace norm of a Hermitian operator, defined as $||A||_1 \equiv\mathrm{Tr}{\sqrt{A^\dagger A}}$ \cite{LogNeg} gives 
	\begin{equation}
	E_{\mathcal{N}}(\rho)\equiv\log_{2}||\rho^{T_{A}}||_1.
	\end{equation}
	Since $\rho^{T_A}=(\rho^{T_B})^T$ and the eigenvalues of matrices $A$ and $A^T$ are the same, $||\rho^{T_{A}}||_1=||\rho^{T_{B}}||_1$, it follows that  $E_{\mathcal{N}}(\rho)$ is uniquely defined for any given bipartition of a given density matrix, $\rho$. Furthermore, a fully separable state can always be written such that the elements $\rho_{ij}=0$, when $i\neq j$ and since $||\rho^{T_A}||_1=1$ meaning that $E_{\mathcal{N}}(\rho)=0$ for any non-entangled state.
	
	When considering multiple parties, one way to classify the entanglement is to look at every bipartite splitting of the system \cite{Dur}. For example, in a four-mode entangled state, $\rho_{1234}$, there are 25 inequivalent measures of entanglement which can be calculated, four of which can be found by considering the splitting $1-234$ and its permutation, three other can be found by looking at the $12-34$ splitting and its permutations, 12 different measures exist when looking at $\mathrm{Tr}(1) 2-34$ and permutations, and finally six other measures are seen when we consider the splitting $\mathrm{Tr}(12) 3-4$ and its permutations. The logarithmic negativity of a splitting like $1-234$ quantifies the strength of the correlations between party $1$ and the three other parties as a whole. Then after observing all splittings, one can say that the multipartite entanglement has increased if the logarithmic negativity has increased in every individual splitting \cite{LogNeg}.
	
	One should note that although these measures are independent of each other since a partial trace is part of local operations and classical communications, the logarithmic negativity under such an operation can only decrease. Meaning that there is an underlying hierarchy in such splittings \cite{LogNeg}. Namely, 
	\begin{equation}
	E_{\mathcal{N}}(\rho_{123-4})\ge E_{\mathcal{N}}(\rho_{\mathrm{Tr}(1)23-4}) \ge E_{\mathcal{N}}(\rho_{\mathrm{Tr}(12)3-4}),
	\end{equation}
	and similarly for other splittings \cite{Dur,LogNeg}. 
	
	\section{Traces}\label{sec:traces}
	In this section, the splittings where certain modes are traced out will be considered. To achieve this extension, Eq. \eqref{eq:phi3mode} also need to be extended to four modes. Hence, using Eq. \eqref{eq:ph-sub} it can be seen that 
	\begin{multline}
		\ket{\phi_1 \left(N,\frac{r}{k+1},\frac{kr}{k+1}\right)}\\
		=\prod_{l=1}^N\hat{S}_l\left(\frac{kr}{k+1}\right)\left((\hat{b}_A\cosh\left(\frac{kr}{k+1}\right)-\hat{b}_A^\dagger\sinh\left(\frac{kr}{k+1}\right)\right)\\
		\times\exp\left(\frac{-r}{2}\left(\frac{\hat{b}_A+\sqrt{n-1}\hat{b}_{B}}{\sqrt{N}}\right.\right.\\
		\left.\left.+\frac{\sqrt{m}\hat{b}_{C}+\sqrt{l}\hat{b}_{D}}{\sqrt{N}}\right)^2-h.c.\right)\ket{vac}_{A,B,C,D},
	\end{multline}
	where $\hat{b}_{B}=\frac{\hat{b}_{i_{1}}+...+\hat{b}_{i_{n}}}{\sqrt{n}}$,  $\hat{b}_{C}=\frac{\hat{b}_{j_1}+...+\hat{b}_{j_m}}{\sqrt{m}}$, and $\hat{b}_{D}=\frac{\hat{b}_{k_1}+...+\hat{b}_{k_p}}{\sqrt{p}}$, such that $n+m+p+1=N$. Also, $i_1,...,i_n$, $j_1,...,j_m$, and $k_1,...,k_p$ are any orderings of the $N-1$ remaining modes once the photon subtracted mode, $A$, has been chosen.
	
	Now the completed version of the plots discussed in the body of this work can be presented. To start off, Fig. \ref{fig:k-N_opt-2} expends upon Fig. \ref{fig:k-N_opt}, and it is by observing the former of these two figures that it can be seen that when $N=8$ and $k\in[0.21,0.33]$ the multipartite entanglement of a four-party scheme is higher than that of a single source, i.e., $k=0$. Similarly, when $N=4$, the logarithmic negativity of the of a single source in the $(AB)_{1/4}-C_{3/4}$ is only surpassed when $k\ge1.02$, while for the $\mathrm{Tr}((AB)_{1/2})C_{1/4}-D_{1/4}$ splitting it is needed that $k\le1.01$. Meaning that there is no situation where the multipartite entanglement is universally better than that of a single source, but significant improvement can be seen in any given splitting using multiple sources.
	
	The whole set of splitting for the Fig. \ref{fig:Nopt} is continued in Fig. \ref{fig:Nopt-2}, which helps complete the general trend that as $k$ decreases, the higher the number of modes, $N$, needs to have an improvement in the four-party multipartite entanglement. 
	
	Finally, Fig. \ref{fig:losses} is completed in \ref{fig:lossesas}. It can be seen in the latter figure that in the $\mathrm{Tr}(D_{1/2}) (AB)_{1/2}-C_{1/4}$ splitting, the loss for the $N=4$, $k=0.82$ curve (blue) decays quite rapidly making its four-party multipartite entanglement more susceptible to loss. However, its bipartite entanglement in the $(AB)_{1/2}-C_{1/2}$ splitting is still at an advantage.

\end{document}